\documentclass{aastex62}
\usepackage{graphicx, amsmath, amssymb, color}
\usepackage{bm}

\begin{document}
\title{Flatness without CMB - the Entanglement of Spatial Curvature and Dark Energy Equation of State}
\author{Haoting Xu}
\affiliation{School of Physics and Astronomy, Sun Yat-sen University, 2 Daxue Road, Tangjia, Zhuhai, China}
\author{Zhiqi Huang}
\affiliation{School of Physics and Astronomy, Sun Yat-sen University, 2 Daxue Road, Tangjia, Zhuhai, China}
\author{Zhenjie Liu}
\affiliation{School of Physics and Astronomy, Sun Yat-sen University, 2 Daxue Road, Tangjia, Zhuhai, China}
\author{Haitao Miao}
\affiliation{School of Physics and Astronomy, Sun Yat-sen University, 2 Daxue Road, Tangjia, Zhuhai, China}

\correspondingauthor{Zhiqi Huang}

\begin{abstract}

The cosmic spatial curvature parameter $\Omega_k$ is constrained, primarily by cosmic microwave background (CMB) data, to be very small. Observations of the cosmic distance ladder and the large scale structure can provide independent checks of the cosmic flatness. Such late-universe constraints on $\Omega_k$, however, are sensitive to the assumptions of the nature of dark energy. For minimally coupled scalar-field models of dark energy, the equation of state $w$ has nontrivial dependence on the cosmic spatial curvature $\Omega_k$~\citep{MH}. Such dependence has not been taken into account in previous studies of future observational projects. In this paper we use the $w$ parameterization proposed by~\cite{MH}, where the dependence of $w$ on $\Omega_k$ is encoded, and perform Fisher forecast on mock data of three benchmark projects: a WFIRST-like Type Ia supernovae survey, an EUCLID-like spectroscopic redshift survey, and an LSST-like photometric redshift survey. We find that the correlation between $\Omega_k$ and $w$ is primarily determined by the data rather than by the theoretical prior. We thus validate the standard approaches of treating $\Omega_k$ and $w$ as independent quantities.
  
\end{abstract}

\section{Introduction \label{sec:intro}}

The plethora of observational data in the past three decades has led to a concordance model of cosmology - a general-relativity-governed universe composed of about 69\% dark energy, 26\% dark matter, and 5\% standard model particles, with small inhomogeneities that originated from vacuum fluctuation during the early-universe inflation. Despite the unknown microscopic nature of dark energy and dark matter, the concordance model has been confronted with, and passed, a host of observational tests - the temperature and polarization anisotropy in cosmic microwave background (CMB) radiation~\citep{COBE, WMAPParam09, PlanckParam15, PlanckParam18}, the type Ia supernovae (SNe) light curves~\citep{Riess98, Perlmutter99, JLA, Pantheon}, the large-scale structure (LSS) of galaxies~\citep{BAO-SDSS-6DF, BAO-SDSS-DR7-MGS, BAO-SDSS-DR12-CMASS, BAO-SDSS-DR12-LOWZ, DESWL18}, and the local measurement of Hubble constant~\citep{Riess18}. The most concise version of the concordance model is the $\Lambda$CDM model, where $\Lambda$ represents the cosmological constant as an interpretation of dark energy, and CDM is the acronym for cold dark matter. 


The early-universe inflation, first proposed to solve the horizon and flatness problems~\citep{Guth1981}, has now become a part of the concordance model. The background spatial curvature of the universe, often characterized by a parameter $\Omega_k$, is predicted to be negligible by most of simple inflationary models. The recently measured temperature and polarization power spectra of cosmic microwave background (CMB), however, give a $99\%$ confidence level detection of a negative $\Omega_k=-0.044^{+0.018}_{-0.015}$, which corresponds to a positive spatial curvature~\citep{PlanckParam18}. The tension between the theoretical expectation and the observation is eased by an addition of CMB lensing reconstruction, which pulls $\Omega_k$ back into consistency with zero to well within $2\sigma$. Further inclusion of baron acoustic oscillations (BAO) data gives a constraint $\Omega_k=0.0007\pm 0.0019$, strongly supporting a spatially flat universe favored by the inflationary paradigm~\citep{PlanckParam18}.

The tension being eased, it is worth noticing that currently there is no fully independent check with comparable accuracy ($\delta \Omega_k \lesssim 0.01$) on the cosmic flatness from other cosmological probes. The constraints on $\Omega_k$ by low-redshift observations, such as SNe, BAO and Hubble constant, often rely on some injection of CMB priors, and are sensitive to the assumptions about the nature of dark energy~\citep{Farooq17, Wang17,Yu18, Joseph18,  Park18}. Strong lensing time delay is a novel tool that in principle can give more model-independent measurements of the cosmic spatial curvature~\citep{Denissenya18}.  Currently available strong lensing data, however, may contain systematic biases that are yet to be understood better~\citep{Li18}.

Assuming a CMB prior and good BAO reconstruction on quasi-nonlinear scales ,\cite{Takada15} show an all-sky, cosmic-variance-limited galaxy survey covering the universe up to $z\gtrsim 4$ can determine $\Omega_k$ to a remarkable accuracy of $\sigma(\Omega_k) \lesssim 10^{-3}$. This forecast can be considered as an ideal limit for future BAO constraints.

In this work we are interested in an explicitly CMB-independent check of the cosmic flatness. More specifically, we consider three experiments that had been proposed - the Wide-Field InfraRed Survey Telescope (WFIRST) supernovae survey, the Euclid spectroscopy redshift survey and the Large Synoptic Survey Telescope (LSST) photometric redshift survey - as our benchmarks. The major configurations of our forecast are taken from the publicly available documents~\citep{WFIRST, Euclid1, LSST}. We do not follow all the details and the recent advances of these projects. (See e.g. \cite{Euclid}.) Thus, we dub the data sets WFIRST-like, Euclid-like and LSST-like, to distinguish between our work and official studies of these projects.

For late-universe observables, $\Omega_k$ has significant degeneracy with dark energy parameters. The standard approach in the literature is to treat $\Omega_k$ and dark energy parameters as independent quantities, and to marginalize over the dark energy parameters. However, this approach is in principle problematic as the evolution of dark energy, and hence its equation of state may depend on the spatial curvature. For instance, if dark energy is quintessence (a minimally coupled canonical scalar field) with a smooth potential~\citep{Wetterich88,Ratra88,Caldwell98,Zlatev99}, its equation of state will depend on the spatial curvature. Such dependence is explicitly calculated in \cite{MH}.

Thus, we use the dark energy parameterization proposed by \cite{HBK} and later improved by \cite{MH}, where the equation of state of dark energy is given by
\begin{equation}
  w = -1 + \frac{2}{3}\left\{\sqrt{\varepsilon_{\phi\infty}} + \left(\sqrt{\varepsilon_s} - \sqrt{\frac{2\varepsilon_{\phi\infty}}{1-\Omega_k}}\right)\left[F\left(\frac{\Omega_ka_{\rm eq}}{\Omega_m}, \frac{a}{a_{\rm eq}}\right) + \zeta_s F_2\left(\frac{a}{a_{\rm eq}}\right)\right]\right\}^2,  \label{eq:3param}
\end{equation}
where $a$ is the scale factor of the universe, normalized to unity today. The pivot $a_{\rm eq}$ is defined as the scale factor at the equality of dark energy and dark matter densities. The ratio of dark matter density to the critical density, $\Omega_m$, enters the formula through its impact on the Hubble drag on the scalar field. The three additional parameters $\varepsilon_s$, $\zeta_s$, and $\varepsilon_{\phi\infty}$ characterize the slope (first derivative) and curvature (second derivative) of the scalar field logarithm potential at the pivot, and the initial velocity of the scalar field, respectively. The functions $F$ and $F_2$, given by
\begin{equation}
  F(\lambda,x) \equiv \frac{3}{x^3} \int^x_0\sqrt{\frac{t^7}{1+\lambda t+t^3}}dt,
\end{equation}
and
\begin{equation}
  F_2(x) \equiv \sqrt{2}\left[1-\frac{\ln\left(1+x^3\right)}{x^3}\right]-\frac{\sqrt{1+x^3}}{x^{3/2}}+\frac{\ln\left[x^{3/2}+\sqrt{1+x^3}\right]}{x^3},
  \end{equation}
can be derived form  dynamic equations of the scalar field. The calculation is tedious but straightforward and can be found in \cite{HBK} and \cite{MH}.

This parameterization contains cosmological constant model ($w=-1$) as a special case (when $\varepsilon_s=\zeta_s = \varepsilon_{\phi\infty}=0$) and covers a wide class of models that can be described by a minimally coupled canonical scalar field. There are models beyond the scope of this parameterization, such as k-essence \citep{Armendariz-Picon00, Armendariz-Picon01}, $f(R)$ gravity \citep{Capozziello03, Carroll04, Nojiri06, Hu07}, etc. The increasing complexity of dark energy model, which we will not cover in this work, may lead to more degeneracy between dark energy parameters and $\Omega_k$.

In addition to the physical parameterization for quintessence model, we also use for comparison purpose a phenomenological dark energy parameterization $w=w_0+w_a(1-a)$ ~\citep{Chevallier01,Linder03}, where the dark energy equation of state does not depend on $\Omega_k$.

This article is organized as follows. In Section~\ref{sec:method} we describe the Fisher forecast method and the mock data. In Section~\ref{sec:results} we give the results and discuss their implications. Section~\ref{sec:conclusions} concludes. Unless otherwise specified, we work with the natural units $c=\hbar=1$.

\section{Fisher Forecast \label{sec:method}}

In this section we give detailed description of the mock data sets and the Fisher forecast technique.

\subsection{WFIRST-like SNe mock data}

The WFIRST-like SNe mock data are generated in seventeen uniform redshift bins spanning a redshift range from $z=0$ to $z=1.7$, with the number of mock samples in each bin listed in Table~\ref{tab:SNe}.

\begin{table}
\caption{Number of supernova samples in each bin for the mock WFIRST-like survey. \label{tab:SNe}}
\begin{center}
\begin{tabular}{crrrrrrrrrrrrrrrrr}
  \hline
  \hline  
$z_{\max}$ & 0.1 & 0.2 & 0.3 & 0.4 & 0.5 & 0.6 & 0.7 & 0.8 & 0.9 & 1.0 & 1.1 & 1.2 
& 1.3 & 1.4 & 1.5 & 1.6 & 1.7 \\
\hline
$N(z)$ & 500 & 69 & 208 & 402 & 223 & 327 & 136 & 136 & 136 & 136 & 136 & 136 & 136 
& 136 & 136 & 136 & 136\\
\hline
\end{tabular}
\end{center}
\end{table}

The distance modulus, namely, the difference between the apparent magnitude $m$ and the absolute magnitude $M$, of a supernova at luminosity distance $d_L$ is given by
\begin{eqnarray}
\mu = 5\log_{10}(\frac{d_L}{\rm Mpc})+25\, . 
\end{eqnarray}
The luminosity distance $d_L$ as a function of redshift $z$ for a given cosmology is calculated with publicly available code CAMB \citep{CAMB}, with minor modification for the dark energy parameterization as done in \cite{HBK} and \cite{MH}. The uncertainty of the distance modulus at redshift $z$ is modeled as
\begin{equation}
  \sigma =\sqrt{\sigma_{\rm meas}^2+\sigma_{\rm int}^2+\sigma_{\rm lens}^2+\sigma_{v}^2},
\end{equation}
where $\sigma_{\rm meas}=0.08$ is the photometric measurement error, $\sigma_{\rm int}=0.09$ is the intrinsic dispersion error, and $\sigma_{\rm lens}=0.07z$ represents gravitational lensing error. Finally, $\sigma_{v} \approx \frac{5\upsilon_{\rm pec}}{\sqrt{2}\ln 10\, cz}$ is due to the redshift uncertainty from the random line-of-sight motion of the sample, assuming a r.m.s. projected peculiar velocity $\upsilon_{\rm pec} = 400\,\mathrm{km\,s^{-1}}$. We ignore possible systematic errors such as correlation between samples, assuming these effects can be properly calibrated. Finally, we marginalize over the absolute magnitude $M$ with a flat prior.

\subsection{LSS mock data}

We consider a spectroscopic Euclid-like redshift survey and a LSST-like photometric redshift survey of galaxies. The galaxy power spectrum is modeled as \citep{Kaiser87, Tegmark97, HVV12, CosmoLib, Chen16, Euclid}
\begin{equation}
P_g(k, \mu; z) = \left(b+ f \mu^2\right)^2 P_m(k) e^{- k^2 \mu^2R_{\parallel}^2 - k^2(1-\mu^2)R_{\perp}^2} + \frac{1}{\epsilon\bar{n}_{\rm obs}}, \label{eq:Pg}
\end{equation}
where $\mu$, not to be confused with the distance modulus discussed in the previous subsection, is the cosine of the angle between the wave vector $\mathbf{k}$ and the line of sight. In the last Poisson noise term on the right-hand side, $\bar{n}_{\rm obs}$ is the number density of observed galaxies, of which a fraction $\epsilon$ of galaxies with successfully measured redshift is used to compute the power spectrum. In Fisher analysis, the wave number vector $\mathbf{k}$ and matter power spectrum $P_m(k)$ are re-calibrated to the reference fiducial cosmology that we use to generate the mock data. 

The linear galaxy bias $b$ is parameterized as
\begin{equation}
  b(z, k) = (b_0+b_1 z)^\beta e^{-\alpha k^2},
\end{equation}
where $b_0$, $b_1$, $\alpha$ and $\beta$ are nuisance parameters. We assume a conservative Gaussian prior $b_0=1\pm 0.05$ to account for the knowledge of galaxy bias and to avoid singularity of Fisher matrix due to perfect degeneracy between bias and the primordial amplitude of density fluctuations. The weak $k$ dependence ($e^{-\alpha k^2}$ factor, with $\alpha \ll \mathrm{Mpc}^{-2}$) allows the baryonic matter to decorrelate with the dark matter on very small scales .

The linear matter power spectrum $P_m(k)$ and the growth $f$, for a given cosmology, again can be computed with CAMB. See e.g. \cite{Euclid} for more details.

The radial smearing scale $R_{\parallel}$ is given by
\begin{equation}
R_{\parallel} = \frac{c\sigma_z}{H},
\end{equation}
where $c$ is the speed of light, $H$ is the Hubble expansion rate at redshift $z$,  and $\sigma_z$ is the combined redshift uncertainty due to the photometric redshift error and the random motion of galaxy along the line of sight. The recipe for $\sigma_z$ is \citep{HVV12, CosmoLib, Chen16}
\begin{equation}
  \sigma_z = (1+z)^2 \sigma_{r0}^2 + \sigma_{\upsilon 0}^2.
\end{equation}
A Gaussian prior is assumed for the nuisance parameter $\sigma_{\upsilon 0}=0.0019\pm 0.0009$. For the uncertainty of redshift measurement, we use $\sigma_{r0} = 0.04$ for photometric redshift (LSST-like mock data) and $\sigma_{r0} = 0.001$ for spectroscopic redshift (Euclid-like mock data).

The smearing in directions perpendicular to the line of sight can be treated either with spherical harmonics or by converting spherical coordinates in a redshift shell to Cartesian coordinates, the latter approach yields, approximately, the transverse smearing scale
\begin{equation}
  R_{\perp} = \frac{c^2\sigma_z\Delta z}{H^2d_{\rm c}},
\end{equation}
where $\Delta z$ is the redshift bin size and $d_{\rm c}$ is  the comoving distance. In the thin-shell limit $R_{\perp}\ll R_{\parallel}$ and the transverse smearing is often ignored in the literature.

We use 30 log-uniform $k$-bins and 30 uniform $\mu$ bins. The result is stable while we increase the number of bins, if we use the following approximate covariance matrix for the galaxy power spectrum in $i$-th bin and $j$-th bin,
\begin{equation}\label{cov}
\mathrm{Cov}\left[P_g(\mathbf{k}_i), P_g(\mathbf{k}_j)\right]= \frac{2\delta_{ij}}{N_i}\left[P_g\left(\mathbf{k}_i\right)\right]^2 + \frac{\sigma_{\min}^2}{\sqrt{N_iN_j}} P_g(\mathbf{k}_i)P_g(\mathbf{k}_j),
\end{equation}
where $N_i$,  the number of independent modes in a survey volume $V_{\rm survey}$ and $i$-th Fourier-space bin with wave number $k_i$ and bin sizes $dk$ and $d\mu$, can be written as
\begin{equation}
  N_i = \frac{(2\pi)^3}{ V_{\rm survey} (2\pi k_i^2 dk d\mu) }.  
\end{equation}

The $\sigma_{\min}$ terms are approximated non-Gaussian corrections to the covariance matrix~\citep{Carron15}. Since a conservative cutoff of quasi-linear scale $k_{\max}$ is used for each redshift bin, the non-Gaussian corrections are expected to be sub-dominant. Thus, we simply use $\sigma_{\min}=1.5\times 10^{-4}$ as estimated in~\cite{Carron15} and ignore the dependence of $\sigma_{\min}$ on survey configurations.

Finally, we summarize the specifications for Euclid-like mock data and LSST-like mock data in Table~\ref{tab:Euclid} and Table~\ref{tab:LSST}, respectively.

\begin{table}
\caption{\small Redshift bins and wavenumber bounds for the Euclid-like mock data. Other fiducial parameters are efficiency  $\epsilon=0.5$, bias $b=\sqrt{1+z}$, sky coverage $f_{\rm sky}=0.36$, spectroscopic redshift error $\sigma_{r0} = 0.001$,  r.m.s. radial motion parameter $\sigma_{\upsilon 0} = 0.0019$. \label{tab:Euclid}}
 \begin{center}
\begin{tabular}{ccccc} \hline\hline
$z_{\rm mean}$ & $z$-range & $\bar{n} [h^{3} {\rm Mpc}^{-3}]$ & $k_{\rm min}[h/$Mpc] & $k_{\rm max}[h/$Mpc]\\
\hline
0.6 & 0.5 -- 0.7 & $3.56 \times 10^{-3}$& 0.0061 & 0.09
\\
0.8 & 0.7 -- 0.9 & $2.82 \times 10^{-3}$& 0.0054 & 0.11
\\
1.0 & 0.9 -- 1.1& $1.81 \times 10^{-3}$& 0.0051 & 0.12
\\
1.2 & 1.1 -- 1.3 & $1.44 \times 10^{-3}$& 0.0048 & 0.14
\\
1.4 & 1.3 -- 1.5& $0.99 \times 10^{-3}$& 0.0047 & 0.16
\\
1.6 & 1.5 -- 1.7 & $0.55 \times 10^{-3}$& 0.0046& 0.18
\\
1.8 & 1.7 -- 1.9 &$ 0.29 \times 10^{-3}$&0.0045 & 0.20
\\
2.0 & 1.9 -- 2.1& $0.15 \times 10^{-3}$ & 0.0045 &0.22
\\
\hline
\end{tabular}
\end{center}
\end{table}

\begin{table}
\caption{Redshift bins and wavenumber bounds for the LSST-like mock data. Other fiducial parameters are efficiency $\epsilon =0.5$; bias $b=1+0.84 z$; sky coverage $f_{\rm sky}=0.58$; photometric redshift error $\sigma_{r0}=0.04$; r.m.s. radial motion parameter $\sigma_{\upsilon 0} = 0.0019$. \label{tab:LSST}}
\begin{center}
\begin{tabular}{ccccc} \hline\hline
$z_{\rm mean}$ & $z$-range & $\bar n [h^{3} {\rm Mpc}^{-3}]$ &$k_{\rm min}[h/$Mpc] & $k_{\rm max}[h/$Mpc]\\
\hline
0.31 & 0.2 -- 0.46 & 0.15& 0.0071 & 0.08
\\
0.55 & 0.46 -- 0.64 &0.10 & 0.0050 & 0.09
\\
0.84 & 0.64 -- 1.04& 0.064 &0.0040 & 0.11
\\
1.18 & 1.04 -- 1.32 &0.036 &0.0035 & 0.14
\\
1.59 & 1.32 -- 1.86&0.017 &0.0030 & 0.17
\\
2.08 & 1.86 -- 2.3 & 0.0069&0.0028& 0.23
\\
2.67 & 2.3 -- 3 & 0.0022 &0.0026 & 0.31
\\
\hline
\end{tabular}
\end{center}
\end{table}

For the cosmological parameters, unless otherwise specified, we use the Planck best-fit $\Lambda$CDM parameters: Hubble constant $H_0=67.32\,\mathrm{km\,s^{-1}Mpc^{-1}}$, fractional matter density $\Omega_m=0.3144$, fractional baryon density $\Omega_b=0.0494$, amplitude and spectral index of primordial power spectrum $A_s = 2.10\times 10^{-9}$, $n_s = 0.966$. Moreover, we assume a minimal neutrino mass $\sum m_{\nu} = 0.06\mathrm{eV}$ and a $0.6\%$ Gaussian prior on $H_0$, which is expected to be achievable by future local distance ladder measurements~\citep{Riess18}.
\section{Results \label{sec:results}}

\begin{figure}
  \centering
  \includegraphics[width=0.48\textwidth]{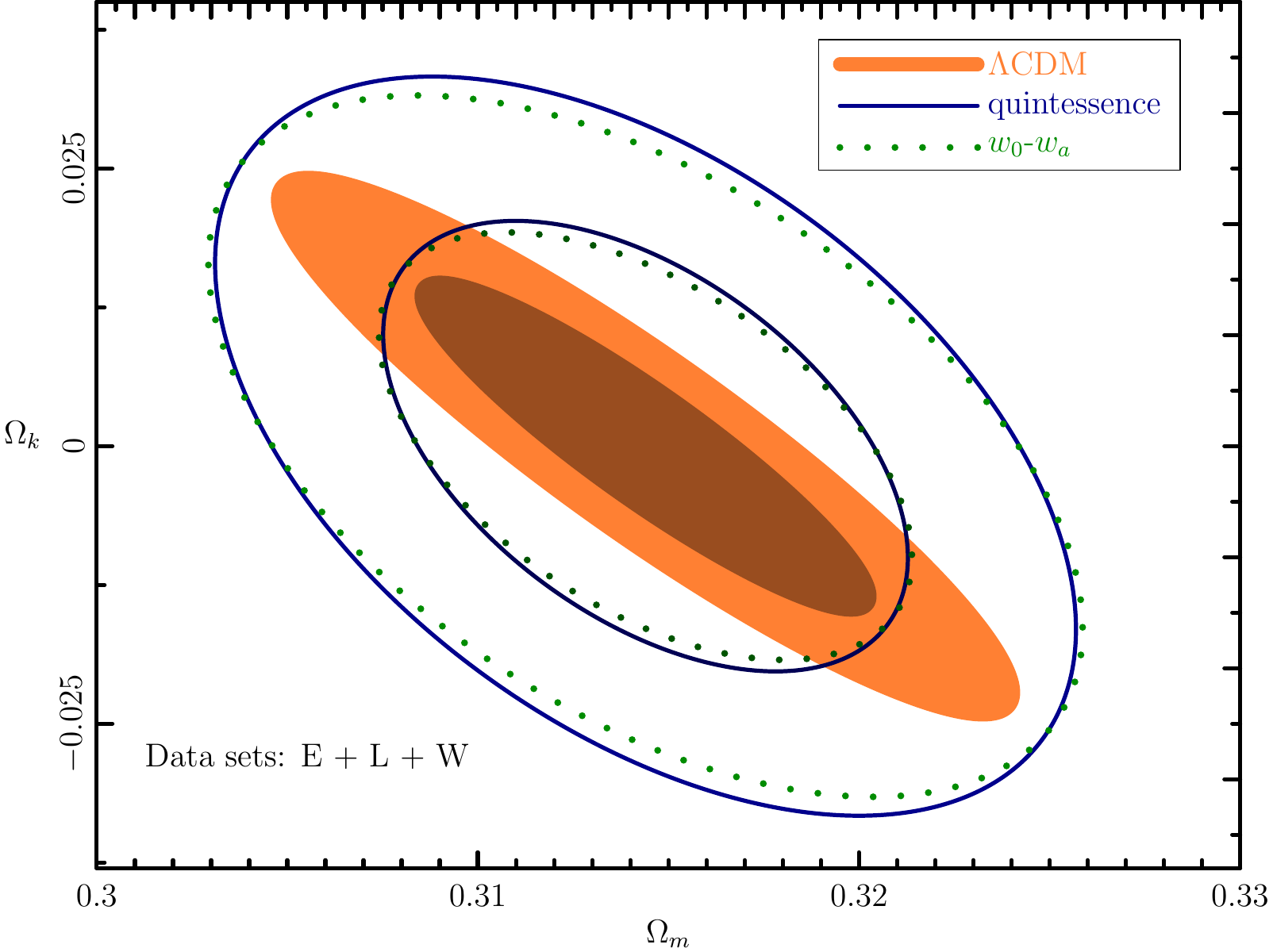} \includegraphics[width=0.48\textwidth]{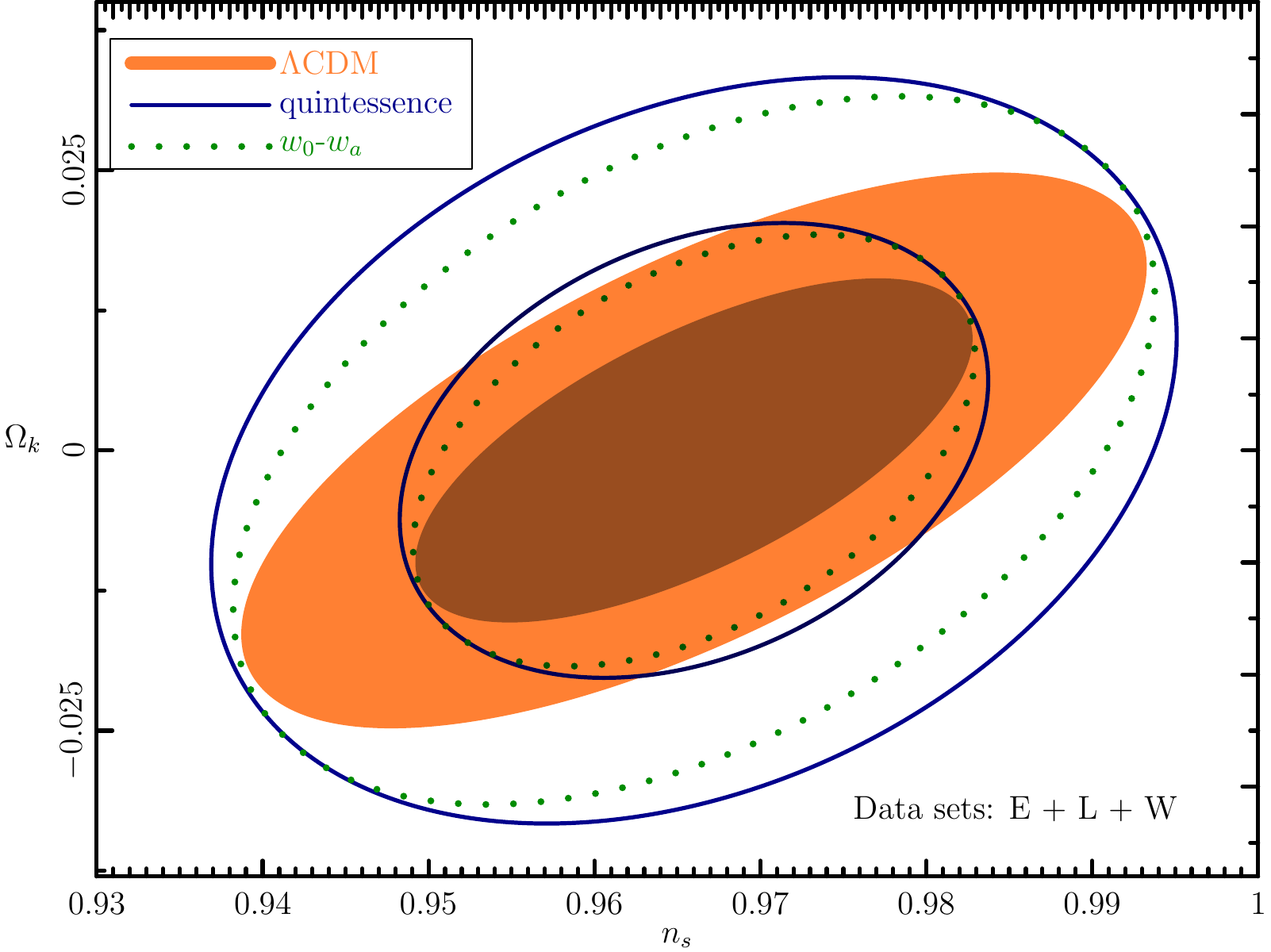}
  \caption{Marginalized constraint in $\Omega_m$-$\Omega_k$ space (left panel) and $n_s$-$\Omega_k$ space (right panel). The inner and outer contours are $1\sigma$ (68.3\% confidence level) and $2\sigma$ (95.4\% confidence level), respectively. All three mock data sets - Euclid-like, LSST-like and WFIRST-like (E+L+W)  are used.\label{fig:ELW}}
\end{figure}

We present the forecast results for non-flat $\Lambda$CDM, quintessence and $w_0$-$w_a$ models in Table~\ref{tab:lcdm}, Table~\ref{tab:qcdm} and Table~\ref{tab:w0wa}, respectively. The constraint on the spatial curvature depends on the dark energy model. However, such dependence becomes very weak while we combine the three mock data sets (WFIRST-like, Euclid-like and LSST-like) together, in which case a percent-level constraint $\sigma(\Omega_k) \approx 0.01$ can be achieved. Comparing the results for quintessence parameterization and for $w_0$-$w_a$ parameterization, we do not find significant difference in the constraints on other parameters, in particular on $\Omega_k$. This statement approximately holds true when we consider combined constraints on multiple parameters. Two examples are given in Figure~\ref{fig:ELW}.

\begin{table}
  \caption{Forecast for (non-flat) $\Lambda$CDM model. Here given are marginalized $1\sigma$ errors of parameters for the Euclid-like redshift survey mock data (E) and the LSST-like redshift survey mock data (L), the WFIRST-like SNe mock data (W), and their combinations. The Planck best-fit $\Lambda$CDM parameters are used~\citep{PlanckParam18}. The fiducial values of five nuisance parameters $b_0$, $b_1$, $\alpha$, $\beta$, $\sigma_{\upsilon 0}$, which we also marginalize over, are given in Table~\ref{tab:Euclid} and \ref{tab:LSST}. Gaussian priors $H_0=67.32\pm 0.4\,\mathrm{km/s/Mpc}$, $b_0=1\pm 0.05$ and $\sigma_{\upsilon 0} = 0.0019\pm 0.0009$ are used. \label{tab:lcdm}}
  \begin{center}
  \begin{tabular}{ccccccc}
    \hline
    \hline
    parameter & W &  E & L &  E+W  & L+W & E+L+W \\
    \hline
    $\Omega_k$          & {\bf 0.036} & {\bf 0.017} & {\bf 0.023} & {\bf 0.015} & {\bf 0.016} & {\bf 0.0097} \\
    $\Omega_m$          & 0.0090      & 0.0074      & 0.0092      & 0.0049       & 0.0062  & 0.0039    \\
    $\Omega_b$          & -           & 0.0018      & 0.0022      & 0.0014       & 0.0019  & 0.0011    \\
    $H_0\,\mathrm{(km\,s^{-1}Mpc^{-1})}$ & -    & 0.36        & 0.35        & 0.34         & 0.34    & 0.24    \\    
    $10^{10}A_s$         & -           & 0.11        & 0.17        & 0.10         & 0.15    & 0.08    \\
    $n_s$               & -           & 0.015       & 0.022       & 0.013        & 0.018   & 0.011    \\
    \hline
  \end{tabular}
  \end{center}
\end{table}

\begin{table}
  \caption{Same as Table~\ref{tab:lcdm}, except for the (non-flat) quintessence model with fiducial $\varepsilon_s=0.3$ and $\varepsilon_{\phi\infty}=\zeta_s = 0$. \label{tab:qcdm}}
  \begin{center}
  \begin{tabular}{ccccccc}
    \hline
    \hline
    parameter & W &  E & L &  E+W  & L+W & E+L+W \\
    \hline
    $\varepsilon_s$     & 13.85 & 1.02 & 0.23 & 0.18 & 0.11 & 0.11 \\
    $\varepsilon_{\phi\infty}$ & 77.11 & 0.51 & 0.20 & 0.37 & 0.17 & 0.15 \\
    $\zeta_s$           &  469.3 & 13.52 & 4.27 & 6.95  & 3.25 & 2.9136 \\
    $\Omega_k$          & {\bf 7.72} & {\bf 0.038} & {\bf 0.037} & {\bf 0.024} & {\bf 0.018} & {\bf 0.013} \\
    $\Omega_m$          & 4.69 & 0.0283 & 0.0151 & 0.0058 & 0.0068 & 0.0045 \\
    $\Omega_b$          & - & 0.0062 & 0.0033 & 0.0015 & 0.0020 & 0.0012 \\
    $H_0\,\mathrm{(km\,s^{-1}Mpc^{-1})}$  
                        & - & 0.38 & 0.38 & 0.36 & 0.36  & 0.25 \\
    $10^{10}A_s$         & - & 0.19 & 0.22 & 0.11 & 0.16 & 0.08 \\
    $n_s$               & - & 0.028 & 0.027 & 0.015 &  0.019 & 0.012 \\
    \hline
  \end{tabular}
  \end{center}
\end{table}

\begin{table}
  \caption{Same as Table~\ref{tab:lcdm}, except for the (non-flat) $w_0$-$w_a$ model with fiducial $w_0=-1$ and $w_a=0.$. \label{tab:w0wa}}
  \begin{center}
  \begin{tabular}{ccccccc}
    \hline
    \hline
    parameter & W &  E & L &  E+W  & L+W & E+L+W \\
    \hline
    $w_0$       & 1.51 & 0.21  & 0.042 & 0.040   & 0.023 & 0.023\\
    $w_a$       & 5.03  & 0.78 & 0.16  & 0.20 & 0.12 & 0.11 \\
    $\Omega_k$  & {\bf 1.07} & {\bf 0.036} & {\bf 0.036} & {\bf 0.020} & {\bf 0.016} & {\bf 0.013} \\
    $\Omega_m$  & 0.43 & 0.0260 & 0.0152 & 0.0058 & 0.0067 & 0.0046 \\
    $\Omega_b$  & - & 0.0053   & 0.0032 & 0.0015 & 0.0021 & 0.0012 \\ 
    $H_0\,\mathrm{(km\,s^{-1}Mpc^{-1})}$  
                &-  & 0.37     & 0.38  & 0.36 & 0.34 & 0.24 \\
    $10^{10}A_s$ & - &  0.20    & 0.23  & 0.11 & 0.16 & 0.08 \\
    $n_s$       & - &  0.031   & 0.028 & 0.014 & 0.018 & 0.011 \\
    \hline
  \end{tabular}
  \end{center}
\end{table}

A cosmological constant does not have dependence on the spatial curvature, whereas a slowly rolling quintessence field does. We have chosen a fiducial $\varepsilon_s = 0.3$, roughly $1\sigma$ bound allowed by current data \citep{MH}, to describe a slowly rolling field that interacts with the spacetime geometry. For a comparison, we switch to a fiducial $\varepsilon_s = 0$ (a very flat quintessence potential) to freeze the field dynamics and to minimize the dependence of $w$ on $\Omega_k$. We find, again, no significant variation of the $1\sigma$ errors on the parameters or of the error contours for multiple parameters.

\section{Conclusions \label{sec:conclusions}}

The combination of upcoming Type Ia supernovae survey and large-volume redshift surveys will confirm (or reject) the cosmic flatness to a remarkable sub-percent precision. Such constraint is not sensitive to the theoretical priors on the connection between dark energy equation of state and the spatial curvature. Thus, we have validated the approaches of treating $\Omega_k$ and $w$ as independent quantities in the literature.

For the forecast of redshift surveys, we have used conservative cutoffs and almost only used information on linear scales. Methods such as BAO reconstruction techniques will provide us with more information on nonlinear scales and improve the constraint on $\Omega_k$ and other parameters~\citep{Takada15}. Alcock-Paczynski effect on nonlinear scales may also be a powerful tool to extract information about the geometry of the universe~\citep{Zhang19}. We leave exploration in these directions as our future work.

\bibliographystyle{aasjournal}

\end{document}